\documentclass[twocolumn,pre,showpacs,superscriptaddress]{revtex4}

\usepackage{graphicx}
\usepackage{amsmath}

\begin{document}

\title{Thermal rectification of a single-wall carbon nanotube: a molecular dynamics study   }

\author{ Azadeh Saeedi $^{1}$, Farrokh Yousefi $^{1}$, Saeed Khadesadr $^{3}$ and M. Ebrahim Foulaadvand $^{1,2}$\\
{\small $^1$ Department of Physics, University of Zanjan, P. O. Box 313, Zanjan, Iran } \\
{\small $^2$ School of Nano-science, Institute for Research in Fundamental Sciences (IPM) , P.O. Box 19395-5531, Teheran, Iran} \\
{\small $^3$ Department of Physics, Tarbiyat Moddares University, P. O. Box 14115-111, Teheran, Iran } \\
 }
\date{\today}

\begin{abstract}

We have investigated the thermal rectification phenomenon in a single-wall mass graded carbon nanotube by molecular dynamics simulation. Second generation Brenner potential has been used to model the inter atomic carbon interaction. Fixed boundary condition has been taken into account. We compare our findings to a previous study by Alaghemandi et al \cite{alaghemandi10} which has been done with a different potential and boundary condition.
The dependence of the rectification factor $R$ on temperature, nanotube diameter and length as well as mass gradient are obtained. It is shown that by increasing the temperature, the rectification decreases whereas by increasing the other parameters namely the mass gradient, diameter and the tube length it increases.

\end{abstract}
\maketitle

\section{introduction}

Rectifcation is a transport process that takes place faster in one
direction than in the opposite one. This phenomenon has attracted
much attention in recent years \cite{roberts}. In electronics this
phenomenon has been used extensively in ubiquitous devices such
as diodes and electric rectifiers \cite{balcerek,jezowski}. The
phenomenon of rectification is not restricted to electronic flow.
In 1970 it was experimentally shown that rectification can occur
in thermal current. For detailed review on thermal rectification
in solid state physics see \cite{dames}. Recently the process of
thermal rectification of heat flow has been detected in nano-sized
materials. It has been empirically shown that externally
mass-loaded carbon and boron nitride nanotubes are capable of
exhibiting thermal rectification \cite{chang}. The empirical
findings of Chang et al have stimulated the interest of
theoreticians on the issue of thermal rectification \cite{chang}.
Thermal property of nanoscale materials is also important both for
fundamental physical theory as well as as for applications
\cite{cahil}. In recent years, different carbon nanotubes (CNT)
\cite{chang,wu,yang08I,yang08II,takahashi} and Graphene Nano
Ribbon (GNR) \cite{hu,yang09,jiang} structures have been proposed
as candidates for thermal rectifers \cite{li04}. These nano-sized
carbon-based materials would have deep implications in thermal
energy control such as on-chip cooling, high efficiency energy
conversion and other phononics applications. Li and co-workers
could theoretically show that thermal rectification appears in a
one-dimensional (1D) chain with non linear interaction among the
masses \cite{li04} and also in a mass-graded monoatomic chain
interacting through the Fermi-Pasta-Ulam (FPU) potential
\cite{yang07,li05}. The anharmonicity is the key feature which
leads the thermal rectification in these one dimensional model
systems. In fact the overlap of the vibrational spectra at the
two ends of the chain is the reason for the existence of the
thermal rectification. Recently Alaghemandi et al have executed
extensive simulation and have shown that mass graded
single-walled carbon nanotubes (SWCNTs) exhibit thermal
rectification \cite{alaghemandi09,alaghemandi10,alaghemandiPRB}.
They have studied the dependence of rectification on the tube
diameter, length, mass gradient and temperature. According to their results,
rectification magnitude increases with increasing the CNT
diameter as well as the mass gradient. Moreover, they
showed that the rectification magnitude decreases when the
temperature is enhanced. They have used reverse non-equilibrium
molecular dynamics \cite{plathe}. The interaction potential
between carbon atoms is adopted from a mechanical viewpoint and
includes radial harmonic, angular and torsion terms \cite{li03}.
In this paper we study the thermal rectification of a SWCNT with
the reactive empirical bond order (REBO) interaction potential
\cite{brener02} between carbon atoms and find the
differences/similarities of our results with those found by
Alagemandi et el.

\section{ Methodology }

We have used classical non equilibrium molecular dynamics (NEMD)
simulation to calculate the thermal conductivity of an armchair
single-wall carbon nanotube (SWCNT). The interaction between two
carbon atoms C-C is modelled by the second-generation reactive
empirical bond order (REBO) potential \cite{brener02}.
Simulations were performed using the LAMMPS package
\cite{lammps}. The velocity Verlet method was employed to
integrate the equations of motion with a time step of one femto
second. First, the entire nanotube is coupled to a Nos\'{e}-Hoover
thermostat at temperature $T$ and MD is performed to equilibrate
and relax the system for 1 ns. After equilibrium, we fix the
atoms of one unit cell (two rings) of carbon atoms from each end
of the CNT. The third unit cell from each end is coupled to a
Nose-Hoover thermostat. The second unit cell is not connected to
a thermostat in order to suppress the phonon reflection from
edges. The hot (cold) reserviour temperature is set to $T+\Delta
T~(T-\Delta T)$ respectively. In our simulations we have taken
$\Delta T=10~K$ . See figure (1) for illustration.

\begin{figure}[h]
\centering
\includegraphics[width=8.5cm,height=10.5cm,angle=0]{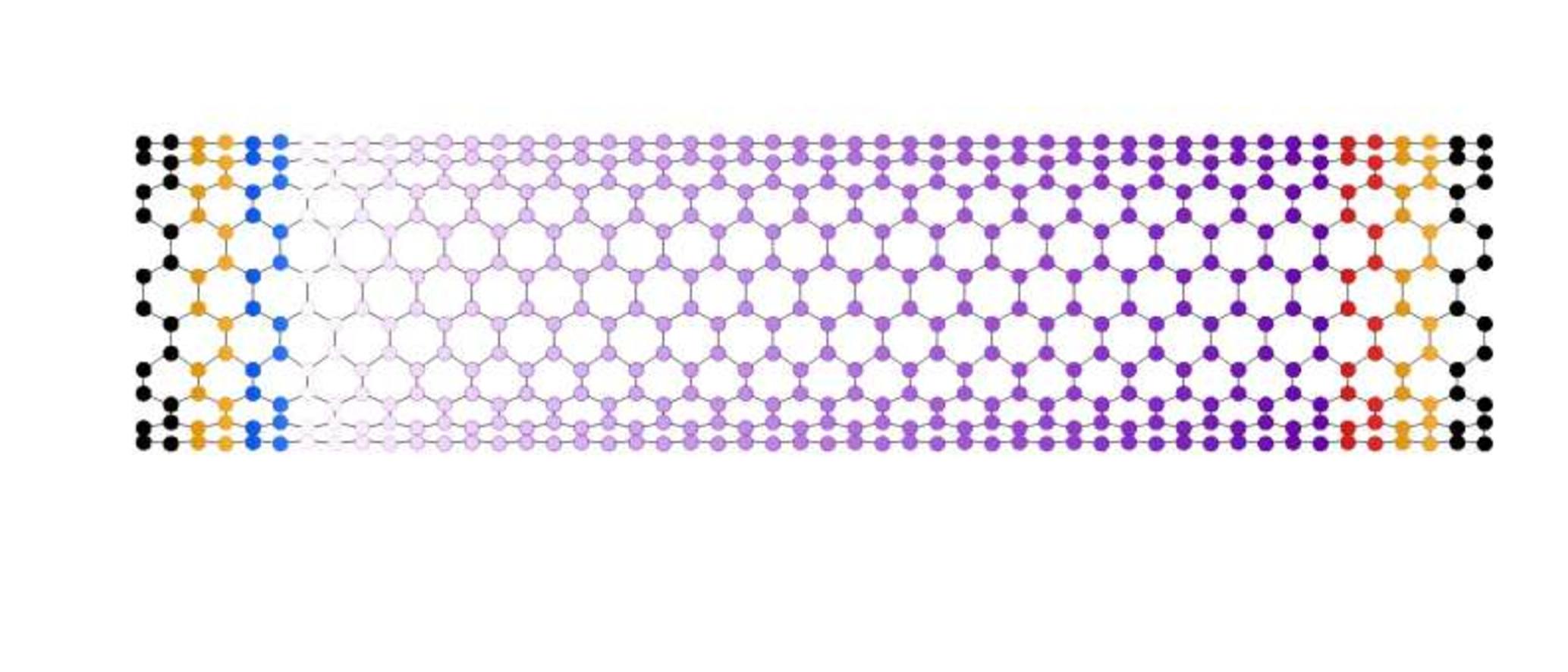}
\caption{Single-walled carbon nanotube under fixed boundary condition. Two layers (black atoms) from each side are kept fixed during the simulation. Blue (red) atoms are connected to the low (high) temperature thermostat. All the other atoms are simulated in the NVE ensemble.
Dark purple indicates high mass region.}\label{Fig1}
\end{figure}

We then perform a run for $10$ ns. As a result of coupling the
system ends to thermostats with different temperatures, a
temperature gradient emerges and a heat flux is established along
the CNT axis. After the system reaches steady state, the thermal
conductivity K of the CNT along its axis is evaluated according to
the Fourier law $Q=-K\frac{dT}{dz}$ as follows:

\begin{eqnarray} K=-\frac{\langle Q(t) \rangle}{dT/dz} \end{eqnarray}

where $\frac{dT}{dz}$ is the temperature gradient along the CNT
axis and the brackets denote time average of thermal flux. We
remark that the instantaneous heat flux ${\bf Q}(t)$ is found from
the following  relation:

\begin{eqnarray} {\bf Q}(t)=\frac{d}{dt}\sum_i {\bf r}_i(t)\epsilon_i(t) \end{eqnarray}

where $\epsilon_i(t)=\frac{p^2_i}{2m} + \frac{1}{2}\sum_{j\neq
i}V({\bf r}_i,{\bf r}_j)$ is the instant total energy of particle
$i$. In computation, people normally do not directly proceed with
(2). Instead, they compute the heat flux in a different manner.
The heat flux is taken as the work done by the thermostat on the
system. In the No$\acute{s}$e-Hoover thermostat \cite{nose} the
equation of motion for particle $i$ becomes:

\begin{eqnarray} \dot{{\bf p}}_i=-\xi{\bf p}_i + {\bf F}_i \end{eqnarray}

where ${\bf F}_i$ is the systematic force exerted on particle
$i$. The heat bath acts on the particle with a force $-\xi{\bf
p}_i$ thus the input power of the heat bath to the system is
$-\xi\frac{{\bf p}_i.{\bf p}_i}{m}$ which can also be regarded as
the thermal current associated to particle $i$. Summation over
$i$ gives:

\begin{eqnarray} {\bf Q}(t)=\sum_i -\xi\frac{{\bf p}_i.{\bf p}_i}{m} \end{eqnarray}

Coming back to (1) the temperature gradient is found by applying a
linear fit to the temperature of the intermediate slabs. The
ratio in Eq. (1) produces a reasonable value provided that the
system is fully equilibrated and the simulation is sampled over a
sufficiently long time. It has been theoretically shown that
implementation of a non uniform mass distribution with a mass
gradient $\alpha$ along the axis of CNT renders possible thermal
rectification \cite{alaghemandi09,alaghemandi10}. The mass of
carbon atoms is assumed to increase with a gradient
$\alpha=\frac{dm}{dz}$ along the CNT axis. This increase mimics
the effect of mass grading by deposition of heavy molecules such
as $C_9H_{16}Pt$ on the CNT \cite{chang}. The left end of the CNT
is set at $z=0$. We have performed extensive MD simulations to
see the differences/similarities with the results of Alaghemandi
et al \cite{alaghemandi10}. The main differences between our
model and reference [17] are as follows: the interatomic
potential in Ref. \cite{alaghemandi10} is adopted from a
continuum mechanics whereas we have used REBO. Second, they used
periodic boundary condition but we have used fixed boundary
condition. We remark that in experiments performed to measure the
thermal conductivity of CNTs, both ends of the CNT is quite kept
fixed (suspended actually over a trench) on a substrate
\cite{yu,choi,li09,hsu,pop} therefore the fixed boundary condition
is more compatible with experiment. Third, they used reverse NEMD
for establishing a thermal current but in our model two
thermostats do the task.
The thermal rectifcation factor $R$ is defined as:\\

\begin{eqnarray} R=\frac{K_{H \rightarrow L}-K_{L \rightarrow H}}{K_{L
\rightarrow H}}\times 100  \end{eqnarray}

Note $K_{H\rightarrow L}$ stands for the thermal conductivity of
the system when heat flows from the high-mass region to the
low-mass region. Visa versa applies to $K_{L \rightarrow H}$.

\section{Results}

In this section we present our simulation results. Figure (2)
shows the rectification factor $R$ versus temperature for a CNT
of length $L=100~ nm$ for two values of chiral indices $(n,m)$
i.e.; $(5,5)$ and $(10,10)$ which correspond to diameters
$d=0.677~nm$ and $d=1.354~nm$ respectively. By increasing the
temperature, the modulus of $R$ decreases towards zero. For low
temperatures, $R$ becomes as large as $-31$ percent. The
difference of $R$ for $n=5$ and $n=10$ becomes smaller when the
temperature is reduced. For example at $T=100~K$ the relative
difference percentage $|\frac{R_5 - R_{10}}{R_{10}}|$ is about
$7$ percent. Our result is qualitatively similar to that obtained
in Ref. \cite{alaghemandi10}. However, our simulations give
larger rectification in comparison to Ref. \cite{alaghemandi10}.
The type of the interatomic potential and the boundary condition
are the most important factors responsible for this difference.

\begin{figure}[h]
\centering
\includegraphics[width=8.5cm,height=6.5cm,angle=0]{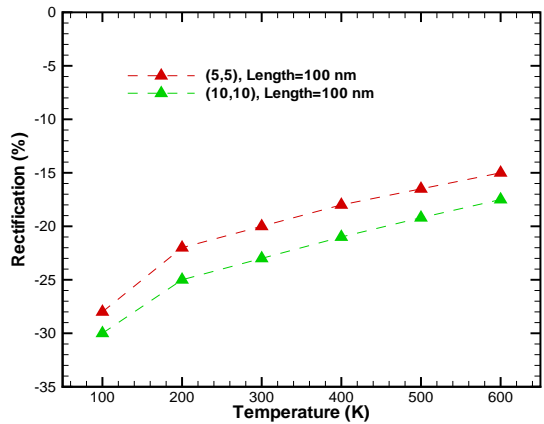}
\caption{Rectification factor versus temperature for armchair $(5,5)$ and $(10,10)$ CNTs having various diameters. Tube length has been $L=100~nm$. Fixed boundary condition has been implemented and $\alpha=5.76~gr/mol~nm$.}\label{Fig1}
\end{figure}

In figure (3) we have shown the temperature dependence of $R$ for
a $(10,10)$ armchair CNT for various lengths from $L=10~nm$ to
$L=100~nm$. The overall behaviour is similar to figure (2). The
magnitude of $R$ decreases with temperature. The larger the tube
length the larger the magnitude of $R$.

\begin{figure}[h]
\centering
\includegraphics[width=8.5cm,height=6.5cm,angle=0]{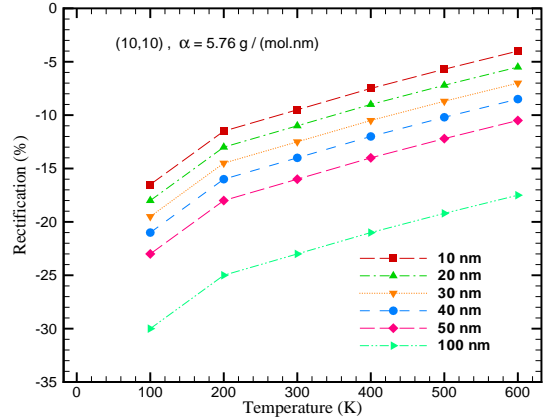}
\caption{Rectification factor versus
temperature for a $(10,10)$ armchair CNT having lengths
$L=10,20,\cdots,100~nm$. Fixed boundary condition has been implemented and $\alpha=5.76~gr/mol~nm$.}\label{Fig1}
\end{figure}

It would be interesting to see how the rectification factor
changes with the variation of mass gradient $\alpha$. Figure (4)
exhibits this dependence for $(5,5)$ and $(10,10)$ CNTs having a
length $L=100~nm$. The temperature has been $T=300~K$. You can see
that by increasing $\alpha$ the rectification factor $R$
increases in magnitude. It goes as high as thirty percent for
$\alpha=13.8~gr/mol~nm$. Moreover, by increasing the CNT diameter
the magnitude of the rectification factor increases. In
\cite{alaghemandi10} we see a qualitative similar behaviour. The
main difference between our results and theirs is that our
simulation shows quite a larger $R$.

\begin{figure}[h]
\centering
\includegraphics[width=8.5cm,height=6.5cm,angle=0]{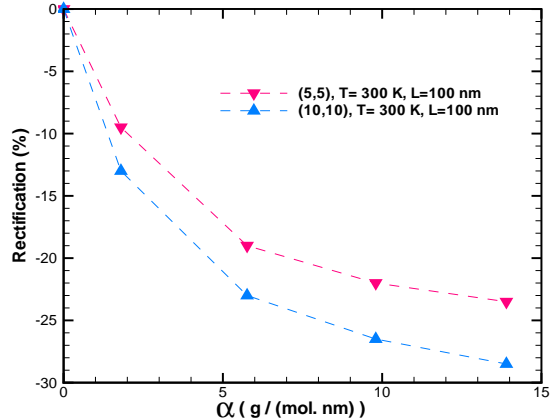}
\caption{Thermal rectification factor as function of the mass gradient $\alpha$ for a $(5,5)$ and $(10,10)$
armchair CNT having lengths $L=100~nm$. The temperature is $300~K$.}\label{Fig1}
\end{figure}

We have also explored the length dependence of the rectification
factor for various diameters. Figure (5) exhibits this
dependence. We observe that by increasing the CNT length, the
absolute value of $R$ increases. It goes as high as twenty four
percent for a $100~nm$ CNT. Moreover, for a fixed length when the
diameter increases the rectification factor $R$ becomes larger in
magnitude.

\begin{figure}[h]
\centering
\includegraphics[width=8.5cm,height=6.5cm,angle=0]{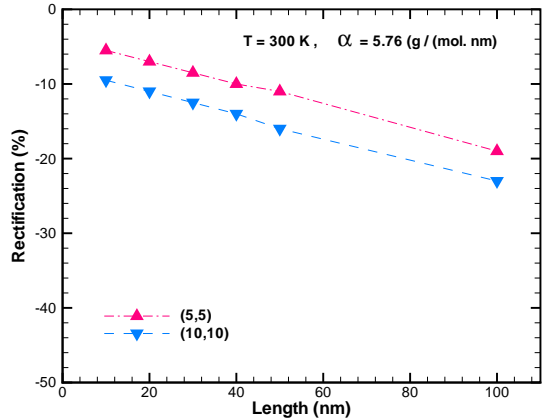}
\caption{Thermal rectification factor as function of CNT length for $(5,5)$ and $(10,10)$ CNTs. The other
parameters are as follows: $\alpha=5.76~gr/mol~nm, T=300~K$.}\label{Fig1}
\end{figure}

Our last figure shows the diameter dependence of the
rectification factor $R$. Figure (6) depicts this behaviour.

\begin{figure}[h]
\centering
\includegraphics[width=8.5cm,height=6.5cm,angle=0]{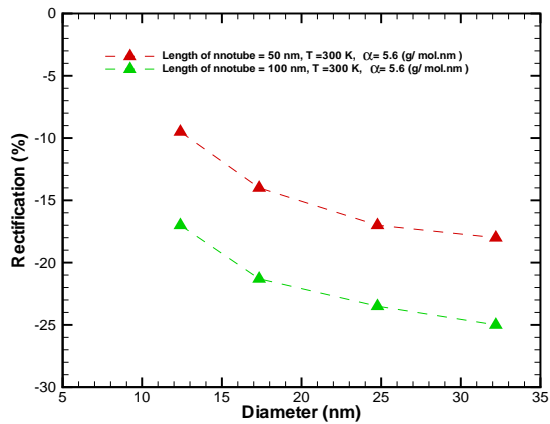}
\caption{Thermal rectification factor as
function of CNT diameter for $L=50,100~nm$ CNTs. The diameters
range from $d=2.16~ nm$ to $d=5.5~nm$. The other parameters are
as follows: $\alpha=5.76~gr/mol~nm, T=300~K$.}\label{Fig1}
\end{figure}

\section{Summary and conclusion}

Thermal rectification in a single-wall mass graded carbon
nanotube is investigated by molecular dynamics simulation. We
compare our findings to a previous study by Alaghemandi et al
\cite{alaghemandi10} which has been done with a different
potential and boundary condition. The dependence of the
rectification factor $R$ on temperature, nanotube diameter and
length as well as mass gradient are obtained. It is shown that by
increasing the temperature, the rectification decreases whereas
by increasing the other parameters it increases in magnitude. Our
results are qualitatively in agreement with reference
\cite{alaghemandi10}. However, there are notable differences
between these two works. Particularly in our findings the
magnitude of rectification factor is larger than those obtained
in \cite{alaghemandi10}. We speculate the type of boundary
condition and phonon scattering from boundaries affects the
problem. Nevertheless, the interatomic potential can play a role
for such difference.

\section{Acknowledgement}

We greatly appreciate the {\it grid centre} of the Institute of
fundamental research in basic sciences (IPM) for providing us
computation facilities. In particular Mr S. Saadatmand, Mr Karimi
and Miss Zeinal Pour. We are thankful to Mohammad Alaghemandi
from technical Darmstadt university, Dr. Ali Naji and Dr. Reza
Asgari from IPM, Dr. Mehdi Vaez Alei from Tehran university and
Dr. Ali Rajabpour from Qazvin university for fruitful help. MEF is
thankful Amir-o Do'leh from Zanjan university and Mehdi sibil
from IPM for useful help.

\end{document}